\documentclass[1p]{elsarticle}
\usepackage{multirow}
\usepackage{tikz}
\usetikzlibrary{positioning}

\makeatletter
\def\ps@pprintTitle{%
     \let\@oddhead\@empty
     \let\@evenhead\@empty
     \def\@oddfoot
     {\hbox to \textwidth%
     {\@myfooterfont%
     }
     }
     \let\@evenfoot\@oddfoot}
\makeatother

\begin{document}

\begin{frontmatter}

\title{Diffusion-based translation between unpaired spontaneous premature neonatal EEG and fetal MEG}

\author[1]{Benoît Brebion}
\ead{benoit.brebion@univ-artois.fr}
\author[2]{Alban Gallard}
\ead{alban.gallard@u-picardie.fr}
\author[3]{Katrin Sippel}
\ead{katrin.sippel@uni-tuebingen.de}
\author[4]{Amer Zaylaa}
\ead{amer.zaylaa@uni-tuebingen.de}
\author[3,4,5]{Hubert Preissl}
\ead{hubert.preissl@uni-tuebingen.de}
\author[2]{Sahar Moghimi}
\ead{sahar.moghimi@u-picardie.fr}
\author[2,6]{Fabrice Wallois}
\ead{fabrice.wallois@u-picardie.fr}
\author[1]{Yaël Frégier}
\ead{yael.fregier@univ-artois.fr}

\affiliation[1]{
    organization={Laboratoire de Mathématiques de Lens (LML), UR 2462, Université d'Artois},
    city={Lens},
    country={France}
}
\affiliation[2]{
    organization={INSERM U1105, Université de Picardie, CURS},
    city={Amiens},
    country={France}
}
\affiliation[3]{
    organization={Department of Internal Medicine IV, University Hospital of Tübingen},
    city={Tübingen},
    country={Germany}
}
\affiliation[4]{
    organization={IDM/fMEG Center of the Helmholtz Center Munich, University of Tübingen, German Center for Diabetes Research (DZD)},
    city={Tübingen},
    country={Germany}
}
\affiliation[5]{
    organization={Institute of Pharmaceutical Sciences, Department of Pharmacy and Biochemistry, Interfaculty Centre for Pharmacogenomics and Pharma Research, Eberhard Karls University Tübingen},
    city={Tübingen},
    country={Germany}
}
\affiliation[6]{
    organization={INSERM U1105, Unit Exploration Fonctionnelles du Système Nerveux Pédiatrique, South University Hospital},
    city={Amiens},
    country={France}
}

\begin{abstract}
Background and objective: Brain activity in premature newborns has traditionally been studied using electroencephalography (EEG), leading to substantial advances in our understanding of early neural development. However, since brain development takes root at the fetal stage, a critical window of this process remains largely unknown. The only technique capable of recording neural activity in the intrauterine environment is fetal magnetoencephalography (fMEG), but this approach presents challenges in terms of data quality and scarcity. Using artificial intelligence, the present research aims to transfer the well-established knowledge from EEG studies to fMEG to improve understanding of prenatal brain development, laying the foundations for better detection and treatment of potential pathologies.
Methods: We developed an unpaired diffusion translation method based on dual diffusion bridges, which notably includes numerical integration improvements to obtain more qualitative results at a lower computational cost. Models were trained on our unpaired dataset of bursts of spontaneous activity from 30 high-resolution premature newborns EEG recordings and 44 fMEG recordings.
Results: We demonstrate that our method achieves significant improvement upon previous results obtained with Generative Adversarial Networks (GANs), by almost 5\% on the mean squared error in the time domain, and completely eliminating the mode collapse problem in the frequency domain, thus achieving near-perfect signal fidelity.
Conclusion: We set a new state of the art in the EEG-fMEG unpaired translation problem, as our developed tool completely paves the way for early brain activity analysis. Overall, we also believe that our method could be reused for other unpaired signal translation applications.
\end{abstract}


\end{frontmatter}

\section{Introduction}

Electroencephalography (EEG) in premature newborns is currently the only possibility to determine direct neural function during early neurodevelopment in humans \cite{BTB_Wallois}. Recently, fetal magnetoencephalography (fMEG) opened up a new avenue of research, as it has the unique ability to record neural activity within the mother womb, providing valuable insights into fetal brain maturation and distinguishing between normal and pathological development \cite{ROTSBS_Preissl, RFEF_Preissl, FMR_Vrba, HFBIM_Vrba, VFAECR_Moraru, QFMA_Vairavan, GDIHFPBA_Linder, FBDGRF_Morin, SDPD_Frohlich}. In particular, the importance of the current work is based on the recent development by switching from superconducting quantum interference devices, which require cooling with liquid helium, to optical pumped magnetometer systems for fetal recordings in utero \cite{FMEGOPM_Corvilain}. This development can lead to an increased interest in human fetal research with biomagnetic recordings.

\subsection{Early brain development}

EEG and event-related potential research on premature infants has shown that brain development during the last trimester of gestation is an extremely dynamic process and that any disturbances in brain wiring can have profound effects on cognitive and sensory-motor skills later in life \cite{DFMS_Bos, SRMA_Fenton, ISFBD_Gilmore, BTB_Wallois, SCCBP_Zmyj}. Before 26 to 28 weeks of gestational age (wGA), the functional maturation of neuronal circuits depends mainly on endogenous mechanisms, partly associated to spontaneous electrical activity. Later, as thalamic afferents move from the subplate to the cortical plate, maturation mechanisms become partially driven by sensory inputs. In addition, the differences between in-utero and ex-utero environments likely impact the communication and organization of neural circuits and structures in both the fetus and premature newborn \cite{AFBNI_Bouyssi, PBLIRO_de_Almeida}. Despite these challenges, studies have demonstrated that both EEG in preterm infants and MEG in fetuses can reveal how immature neural circuits process sensory stimuli, such as sounds and visual cues \cite{VFERF_McCubbin, DAEFHF_Holst}. In this direction, rare indications also suggest relatively similar signatures of spontaneous neural activity in the EEG of premature newborns and the MEG of fetuses at the same gestational age \cite{NIDIBAP_Eswaran}.

It is also acknowledged that the resting state activity of early developing neural networks is characterized by discontinuity, with bursts of activity separated by quiescent periods, known as inter-burst intervals (IBIs), and lasting a few seconds \cite{GFMA_Vairavan, BTB_Wallois, EBA_Luhmann}. During these bursts, specific EEG graphoelements linked to different stages of gestation can be observed: theta temporal activity in coalescence with slow-waves (TTA-SW) between 24 and 32 wGA, delta brushes (DB) between 28 and 36 wGA, and frontal transients (FT) between 36 and 42 wGA \cite{RGTCE_Kane, BTB_Wallois, NEEG_BourelPonchel}. These neurobiomarkers are associated with spontaneous neural activities occurring at both cortical and subcortical levels \cite{PNNN_Routier, EPADC_Khazipov, FSTN_Routier}, and could be modulated by the environment \cite{CAERPN_Kaminska, IRBCG_Moghimi}. Excessively long discontinuity of the IBIs at a given age, as well as the absence of the aforementioned EEG graphoelements can be indicative of neurological outcomes \cite{IPN_Wallois}. Some of these neural features, such as discontinuities, have also been observed in fetal MEG \cite{NIDIBAP_Eswaran}. Although there are established automatic techniques for detecting spontaneous neural activity \cite{ADFBS_Moser, DDPSBA_Vairavan}, there is a lack of comprehensive studies defining fetal manifestations of spontaneous activity, meaning brain activity patterns of the fetus are still mostly unknown.

\subsection{Unpaired data translation}

To address this, we build upon the hypothesis proposed in \cite{TSPN_Gallard} that EEG signals can be effectively translated into fMEG signals of similar gestational ages using Deep Learning, providing valuable insights into the manifestation and detection of neurobiomarkers of the fetus. But due to the infeasibility of simultaneous fetal EEG and MEG recordings, one must use unpaired datasets. The concept of unpaired translation models has gained significant interest over the years, especially since the introduction of the CycleGAN method \cite{CycleGAN_Zhu}. These models allow for the mapping between two domains without requiring paired examples, which is particularly useful in a context like ours where such paired data is non-existent. Initially, this approach was used in the medical field for image-to-image translation \cite{MedGAN_Armanious, CycleMedGAN_Armanious}, and it has since been adapted for time series applications as well \cite{FetalECG_Mohebbian, CardioGAN_Sarkar, MTSCycleGAN_Schockaert}. In particular, the problem of EEG-fMEG unpaired translation has been addressed from this point of view by \cite{TSPN_Gallard}. However, by relying on generative adversarial networks (GANs) as their backbone, these models share a well known inherent issue called mode collapse \cite{MCGAN_Kossale}, where the generator produces a limited variety of outputs despite the potential for a much broader spectrum. This phenomenon thus hinders the model's ability to generalize effectively across diverse aspects of the data.

In recent years, diffusion models \cite{DULNT_SohlDickstein, GMEGDD_Song, DDPM_Ho, SBGM_Song} have emerged as the state-of-the-art for generative applications, offering a robust alternative to GANs \cite{IDDPM_Nichol, DMBG_Dhariwal}. These models progressively transform noise into structured data, effectively avoiding the mode collapse problem. Despite their advantages, the application of diffusion models has predominantly focused on images, which remains the main explored area in generative research \cite{ZSTIG_Ramesh, LDM_Rombach, PTIDM_Saharia}. One can note that diffusion models have also been used for time series in the literature, but only for a few main tasks: generation \cite{DiffWave_Kong, DBCECG_Alcaraz, OCTSGP_Coletta}, forecasting \cite{ScoreGrad_Yan, TimeGrad_Rasul, D3VAE_Li, TDSTF_Chang}, imputation \cite{CSDI_Tashiro, SSSD_Alcaraz}, anomaly detection \cite{ImDiffusion_Chen, DDMT_Yang}, and noise removal \cite{DeScoDECG_Li}. This image-centric focus is even more pronounced when considering unpaired diffusion translation models \cite{UNITDDPM_Sasaki, CycleDiffusion_Wu, EGSDE_Zhao, UMIT_Ozbey, DDIB_Su}, where studies have yet to really extend their application to time series. It is worth mentioning that a few diffusion-based translation models for biosignals exists, but these models always require paired data \cite{RDDM_Shome, PPG2ECG_Belhasin}. These signal-processing applications are still in their infancy, and there is a significant gap in research exploring the unpaired translation of complex neurophysiological signals like EEG and fMEG.

\subsection{Contributions}

In this article, we introduce a diffusion-based unpaired EEG-fMEG translation technique that greatly improves previous results obtained in \cite{TSPN_Gallard} with CycleGAN, while eliminating mode collapse problems. By enabling the translation of preterm infants EEG signals into fetal MEG, our method can help clinicians better understand the manifestations of neural activity across different stages of early brain development. This could ultimately contribute to the monitoring of brain evolution and early signs of neuropathologies in the fetus, in order to subsequently provide appropriate treatment.

We also demonstrate, through an ablation study, the relevance of the improvements we add to the initial diffusion backbone method. Finally, we provide the source code of our pipeline, to encourage adaptations to other unpaired time series translation problems.

\section{Related work}

\subsection{CycleGAN}

CycleGAN \cite{CycleGAN_Zhu} consists of two maps $f: X \rightarrow Y $ and $g: Y \rightarrow X$ between two domains $X$ and $Y$ that are trained as conditional generators: $f$ (resp. $g$) is seen as a generator for the domain $Y$ (resp. $X$), conditioned on elements of the domain $X$ (resp. $Y$) with GAN losses, together with a cycle consistency loss of the form:
\begin{equation}
    \| f \circ g - Id_{Y}\|^2 + \| g \circ f - Id_{X}\|^2,
    \label{cycle}
\end{equation}
that aims to minimize the error between the original data and the reconstructed data after translations in both directions.

\subsection{Diffusion models}

Diffusion models have emerged as a powerful generative modeling framework by gradually transforming a simple noise distribution into a complex data distribution through a series of reversible steps.

Denoising Diffusion Probabilistic Models (DDPM) \cite{DDPM_Ho} define a forward noising process as a Markov chain which progressively corrupts a data sample \( \mathbf{x}_0 \) by adding Gaussian noise over \( T \) timesteps. After sufficient steps, the sample \( \mathbf{x}_T \) approaches an isotropic Gaussian distribution. Each transition of this process can be described by the conditional distribution:
\begin{equation}
    q(\mathbf{x}_t | \mathbf{x}_{t-1}) = \mathcal{N}(\mathbf{x}_t ; \sqrt{\alpha_t} \mathbf{x}_{t-1}, (1 - \alpha_t)\mathbf{I}),
\end{equation}
where \( \alpha_t \) controls the noise variance at each step \( t \). Then, a reverse denoising process is defined by a learnable Markov chain of the form:
\begin{equation}
    p_\theta(\mathbf{x}_{t-1}|\mathbf{x}_t) = \mathcal{N}(\mathbf{x}_{t-1} ; \mu_\theta(\mathbf{x}_t,t), \Sigma_\theta(\mathbf{x}_t,t)),
\end{equation}
which incrementally denoises \( \mathbf{x}_T \) back to \( \mathbf{x}_0 \) by parameterizing the reverse distribution as a Gaussian whose mean $\mu_\theta$ and variance $\Sigma_\theta$ are predicted by a model with parameters $\theta$. The model is generally trained by minimizing a simplified objective that directly estimates the added noise.

Score-based diffusion using Stochastic Differential Equations \cite{SBGM_Song} (which we refer to as Score SDE) can be seen as a generalization of DDPM from a discrete to a continuous diffusion process $\{\mathbf{x}(t)\}_{t=0}^T$, where $\mathbf{x}(0) \sim p_0$ and $\mathbf{x}(T) \sim p_T$ with $p_0$ the data distribution and $p_T$ the prior Gaussian distribution. The forward diffusion process is thus defined using a Stochastic Differential Equation (SDE):
\begin{equation}
    \mathrm{d}\mathbf{x} = \mathbf{f}(\mathbf{x},t)\mathrm{d}t + g(t)\mathrm{d}\mathbf{w},
\end{equation}
where $\mathbf{f}(\mathbf{x},t)$ and $g(t)$ are respectively the drift and diffusion coefficients of the SDE, and $\mathbf{w}$ is a standard Wiener process (Brownian motion). This process can then be reversed by solving the following reverse-time SDE \cite{RTDEM_Anderson}:
\begin{equation}
    \mathrm{d}\mathbf{x} = \left[ \mathbf{f}(\mathbf{x},t) - g(t)^2 \nabla_\mathbf{x} \log p_t(\mathbf{x}) \right] \mathrm{d}t + g(t)\mathrm{d}\bar{\mathbf{w}},
\end{equation}
where $\bar{\mathbf{w}}$ is a standard Wiener process running backwards in time (from $T$ to $0$), and the score function $\nabla_\mathbf{x} \log p_t(\mathbf{x})$ is approximated by training a time-dependent neural network $\mathbf{s}_\theta(\mathbf{x},t)$. This reverse SDE can also be converted to a deterministic Probability Flow Ordinary Differential Equation (PF-ODE):
\begin{equation}
    \mathrm{d}\mathbf{x} = \left[ \mathbf{f}(\mathbf{x},t) - \frac{1}{2}g(t)^2 \nabla_\mathbf{x} \log p_t(\mathbf{x}) \right] \mathrm{d}t,
    \label{score_sde_pf_ode}
\end{equation}
which shares the same marginal distributions as the SDE. The trained network can be seen as a vector field, and data generation is thus performed by numerically solving either the reverse SDE or its equivalent PF-ODE.

Alongside Score SDE, Denoising Diffusion Implicit Models (DDIM) \cite{DDIM_Song} introduce a non-Markovian design derived from DDPM, which allows for a faster and more flexible sampling process that can skip intermediate steps. This is achieved by defining a deterministic sampling schedule, which can be viewed as a particular discretization of an underlying PF-ODE.

\subsection{Dual Diffusion Implicit Bridges}

Dual Diffusion Implicit Bridges (DDIB) \cite{DDIB_Su} is a popular technique developed for unpaired image-to-image translation based on DDIM \cite{DDIM_Song} and diffusion Schrödinger bridges \cite{SB_Schrodinger, DSB_Bortoli}.

This approach relies on the individual training of two diffusion models $v_\theta^{(s)}$ and $v_\theta^{(t)}$ on the source $s$ and target $t$ domains. Then, at inference, translation between those domains is done with a two step process. First, a forward PF-ODE is integrated using a numerical method $\mathrm{ODESolve}$, inverting a sample $\mathbf{x}^{(s)}$ from the source distribution to a latent version $\mathbf{x}^{(l)}$ from the prior Gaussian distribution:
\begin{equation}
    \mathbf{x}^{(l)} = \mathrm{ODESolve}(\mathbf{x}^{(s)} ; v_\theta^{(s)}, 0, T).
    \label{forward_ddib}
\end{equation}
The integration endpoints $0$ and $T$ correspond to the noise schedule: starting from clean data at $t=0$ and progressing to fully corrupted data at $t=T$. Then, a reverse PF-ODE is also solved, producing $\mathbf{x}^{(t)}$, the target distribution counterpart of $\mathbf{x}^{(s)}$, from $\mathbf{x}^{(l)}$:
\begin{equation}
    \mathbf{x}^{(t)} = \mathrm{ODESolve}(\mathbf{x}^{(l)} ; v_\theta^{(t)}, T, 0).
    \label{reverse_ddib}
\end{equation}
This process can also be performed from the target to the source distribution by inverting $t$ and $s$ in (\ref{forward_ddib}) and (\ref{reverse_ddib}). When applied consecutively, i.e. from source $\mathbf{x}^{(s)}$ to latent $\mathbf{x}^{(l)}$ to target $\mathbf{x}^{(t)}$ and then back to latent $\mathbf{x}'^{(l)}$ to source $\mathbf{x}'^{(s)}$, cycle consistency $\mathbf{x}^{(s)} = \mathbf{x}'^{(s)}$ is guaranteed, up to numerical errors from the ODE solvers.

To draw a parallel with Schrödinger Bridges, the innovation of DDIB lies in constructing an implicit bridge between two diffusion models, that have been separately trained on their respective domain. The term bridge here refers to a shared latent space that connects the noisy distributions of both models diffusion processes. During inference, an input sample from the source modality is mapped to a noisy representation via its forward diffusion process. This noisy representation is then interpreted as lying within the shared stochastic bridge and passed through the reverse diffusion process of the target data model, generating the translated sample. Because both models noise trajectories are implicitly aligned, the resulting output is coherent and semantically consistent with the source input. Crucially, this method avoids generating incorrect translations by ensuring that the transition through the shared noise space preserves the structural and semantic constraints learned by each model. In essence, the diffusion models provide a mechanism for finding this most probable path from source to target data.

\subsection{Unpaired diffusion-based translation methods}

Other avenues have also been explored for unpaired diffusion-based translation models. UNIT-DDPM \cite{UNITDDPM_Sasaki} relies on a joint training of 2 DDPMs and a cycle consistency loss for translation. In the scope of text-conditioned diffusion models, CycleNet \cite{CycleNet_Xu} also employs a cycle consistency regularization. CycleDiffusion \cite{CycleDiffusion_Wu} and EGSDE \cite{EGSDE_Zhao} are extensions of DDIB working with stochastic differential equations. SynDiff \cite{UMIT_Ozbey} uses a cycle-consistent architecture and an adversarial projector inspired by GANs for reverse diffusion sampling. UNSB \cite{UNSB_Kim} relies on Schrödinger
Bridges but with a specific focus on high-resolution image translation. Finally, IBCD \cite{IBCD_Lee} recently proposed the use of distillation with DDIB to reduce the number of model steps.

\section{Method}

\subsection{Dual Diffusion Implicit Bridges extended with Elucidated Diffusion Models}

In the original paper \cite{DDIB_Su}, Dual Diffusion Implicit Bridges (DDIB) based their method on the parameterization proposed by Denoising Diffusion Implicit Models (DDIM) \cite{DDIM_Song}. While allowing both a forward and reverse deterministic diffusion process, and being a major step towards faster diffusion models, DDIM is essentially restrained to a probabilistic ODE integrated with a first-order numerical solver.

We instead propose to use the PF-ODE introduced in Elucidated Diffusion Models (EDM) \cite{EDM_Karras}:
\begin{equation}
    \mathrm{d}\mathbf{x} = \left[ \frac{\dot{s}(t)}{s(t)} \mathbf{x} - s(t)^2 \dot{\sigma}(t) \sigma(t) \nabla_\mathbf{x} \log p \left( \frac{\mathbf{x}}{s(t)} ; \sigma(t) \right) \right] \mathrm{d}t,
\end{equation}
where $\sigma(t)$ and $s(t)$ are respectively the time and scale schedules, and the dot corresponds to a time derivative. This formulation relies on the fact that the objective of the PF-ODE is to match a particular set of marginal distributions. As a result, it disposes of the functions $\mathbf{f}$ and $g$ present in the original PF-ODE (\ref{score_sde_pf_ode}), and from which the properties of the marginal distributions could only be deduced indirectly. The diffusion framework introduced in EDM also allows the use of higher-order solvers, unlike DDIM. Consequently, through extensive experiments, they show that Heun's 2nd order method \cite{Heun_Ascher} is a great trade-off between sampling speed and discretization error, surpassing first-order ODE solvers in nearly all cases. Thus, we seek to incorporate this solver into DDIB, a potential improvement which was already mentioned in the original paper \cite{DDIB_Su} and was recently explored in the audio domain \cite{LDBATT_Mancusi}. We are also adapting the method accordingly, and in particular the model used, in order to be applied with signals instead of images.

\begin{figure}
    \centering
    \resizebox{\columnwidth}{!}{
        \begin{tikzpicture}
            \node[draw, on grid, inner sep=2mm, rounded corners, line width=0.5mm, draw=blue] (eeg) {EEG: $\mathbf{x}^{(s)}$};
            \node[draw, on grid, inner sep=2mm, rounded corners, line width=0.5mm, draw=blue, right=4cm of eeg] (noised_eeg) {Noised EEG: $\mathbf{x}^{(l)}$};
            \node[draw, on grid, inner sep=2mm, rounded corners, line width=0.5mm, draw=red, right=4cm of noised_eeg] (transl_meg) {Transl. fMEG: $\mathbf{x}^{(t)}$};
            \node[draw, on grid, inner sep=2mm, rounded corners, line width=0.5mm, draw=red, below=2cm of transl_meg] (noised_transl_meg) {Noised transl. fMEG: $\mathbf{x}'^{(l)}$};
            \node[draw, on grid, inner sep=2mm, rounded corners, line width=0.5mm, draw=blue, below=2cm of eeg] (reconst_eeg) {Reconst. EEG: $\mathbf{x}'^{(s)}$};
            
            \draw[->,>=latex, line width=0.3mm] (eeg) edge [out=45, in=135] node [above=0.1cm, pos=0.4, align=center] (arrow_1) {$\mathrm{ODESolve}(\mathbf{x}^{(s)} ; v_\theta^{(s)}, 0, T)$} (noised_eeg);
            \draw[->,>=latex, line width=0.3mm] (noised_eeg) edge [out=45, in=135] node [above=0.1cm, pos=0.6, align=center] (arrow_2) {$\mathrm{ODESolve}(\mathbf{x}^{(l)} ; v_\theta^{(t)}, T, 0)$} (transl_meg);
            \draw[->,>=latex, line width=0.3mm] (transl_meg) edge [out=-45, in=45] node [right=0.1cm, midway, align=center] (arrow_3) {$\mathrm{ODESolve}(\mathbf{x}^{(t)} ; v_\theta^{(t)}, 0, T)$}(noised_transl_meg);
            \draw[->,>=latex, line width=0.3mm] (noised_transl_meg) edge [out=-135, in=-45] node [below=0.1cm, midway, align=center] (arrow_4) {$\mathrm{ODESolve}(\mathbf{x}'^{(l)} ; v_\theta^{(s)}, T, 0)$} (reconst_eeg);
        \end{tikzpicture}
    }
    \caption{Overview of our diffusion-based translation method at inference using the previously trained models $v_\theta^{(s)}$ and $v_\theta^{(t)}$, here from a source EEG to a target fMEG signal. The upper part corresponds to the translation of the signal $\mathbf{x}^{(s)}$ into its counterpart $\mathbf{x}^{(t)}$, while the lower part is optional and transforms the signal back into its original type $\mathbf{x}'^{(s)}$, with cycle consistency $\mathbf{x}^{(s)} = \mathbf{x}'^{(s)}$ in theory.}
    \label{overview_ddib_eeg_meg}
\end{figure}
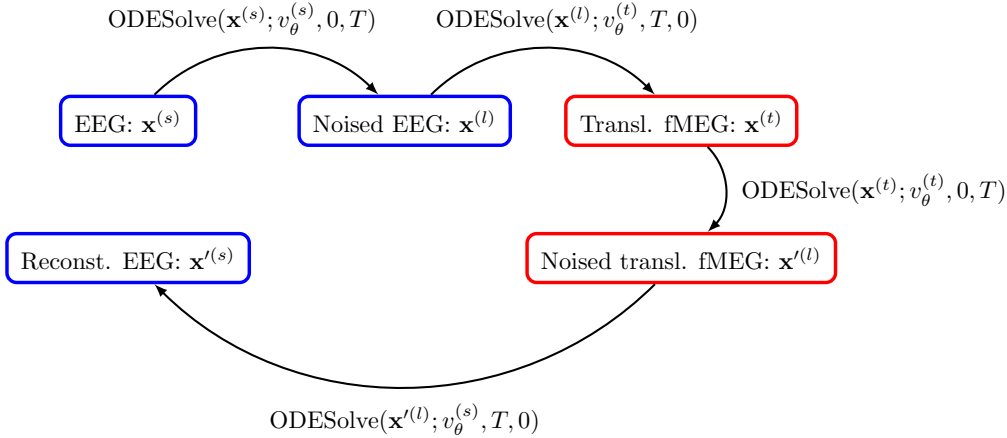

To summarize, our method is a 1D-adapted version of DDIB using EDM as a backbone instead of DDIM, enabling us to use a more efficient solver than the original version, and which we then apply on our EEG-fMEG problem. An overview is shown in Figure \ref{overview_ddib_eeg_meg}.

\subsection{Architecture, training, and sampling details}

Our complete pipeline is written in PyTorch. The network architecture used in this study is the U-Net introduced in DDPM \cite{DDPM_Ho}, adapted for 1D signals. The EDM framework is utilized as the overall training mechanism, with Heun's method solver corresponding to their Algorithm 1 \cite{EDM_Karras}. As in DDIB, 2 models are trained separately on each data type, namely EEG and fMEG. Training was conducted using 1x NVIDIA RTX 8000 (48GB) GPU for around 120 epochs (30.000 iterations with a batch size of 32), although VRAM needs don't exceed 6GB. It took around 1 and a half hour of clock time to train each model with our settings.

\section{Evaluation}

\subsection{Dataset}

\textit{Additional details about dataset acquisition and processing are available in Supplementary Material.}

In this work, we used high-resolution EEG recordings of 30 healthy sleeping premature neonates aged between 34 and 37 weeks gestational age (mean age 35.47 ± 1.09 wGA) collected at Amiens University Hospital's neonatal intensive care unit (Amiens, France), with recordings lasting from 6.5 to 42.9 minutes (average 22.9 ± 9.4 minutes). A zero-phase band-pass filter (0.5-20 Hz) was applied, corresponding to the frequencies of cortical manifestations of spontaneous activity at this period of neurodevelopment \cite{BTB_Wallois}, and the spatial resolution was reduced from 64/128 channels to a low-resolution 10 bipolar-channel montage, commonly used in clinical settings to match the number of MEG channels reflecting brain activity \cite{OEEGEPM_Sazgar, ACNSS_Koolen, RAEEG_Stevenson}. Artifacts exceeding an amplitude threshold of 500 $\mu$V were removed, along with a 0.5-second window before and after each one detected.

We also utilized fMEG recordings of 44 fetuses aged between 34 and 37 wGA (mean age 35.34 ± 1.10 wGA) from Tübingen Hospital (Tübingen, Germany), with recordings ranging from 6 to 27 minutes (average 22.1 ± 8.1 minutes). Cardiac artifacts that interfered with the measurement of brain activity were removed using the orthogonal projection method \cite{FMEGRPO_Vrba, ORMCGFMEG_McCubbin}. This method was applied twice consecutively given the presence of two distinct interfering components, namely maternal and fetal cardiac activity. The number of projectors used in each step was manually selected (1 to 3 for each phase) for optimal performance assessed through visual inspection, in order to preserve background neural activity. A zero-phase band-pass filter (0.5-20 Hz) was also applied, followed by visual inspection to eliminate any noisy artifacts that could mislead the burst-detection phase. Remaining artifacts were removed using a 2000 fT amplitude threshold, along with a 0.5-second window before and after each artifact detected.

To automatically detect bursts in premature EEG and fMEG recordings and retain only signals containing activity, a nonlinear energy operator (NLEO) was used. It characterizes bursts by combining the signal's amplitude and spectral content at each time point \cite{ONLEOEEG_Palmu, DEEGB_Palmu}. To better capture dynamic changes, signals were resampled at 256 Hz, and the absolute value of the NLEO index was smoothed by calculating the average over a sliding window of 384 samples (equivalent of 1.5 s) \cite{DEEGB_Palmu}. A threshold was then applied to each channel, with signals above the threshold identified as bursts and those below as inter-burst intervals (IBIs). This threshold was manually optimized for each subject through visual inspection, with mean values for EEG and fMEG signals of 0.81 ± 0.23 $\mu$V² and 17.0 ± 6.4 fT² respectively. Based on previous reports \cite{DBSDMP_Segovia, MEGSCPB_Moser}, and the estimated fetal head diameter at this gestational age (around 10 cm) \cite{CFSHD_Chitty}, we anticipated that spontaneous neural activity would be detected in only a limited number of MEG channels. Accordingly, a 10-channel region of interest was defined on the basis of the spatial amplitude distribution of smoothed fMEG NLEO signals. This region was centered around the channel with the maximum amplitude, and the nine nearest channels were selected to obtain the 10 fMEG channels likely to present signals corresponding to neural activity. A burst was then marked if it occurred simultaneously across at least half of the channels, and otherwise the period was labeled as an inter-burst interval (IBI). Bursts separated by an IBI shorter than 2 seconds were merged, and only burst periods were considered for further analysis.

Finally, to reduce the number of features provided as model inputs while maintaining data quality, EEG and fMEG bursts were resampled at 64 Hz and divided into 5-second segments with a 2.5-second overlap (bursts shorter than 5 seconds were excluded). One must note that we work with single-channel data, meaning each channel of a segment is considered as an individual burst. We also apply a normalization of range [-1,1] based on the minimum and maximum values of the entire segment dataset. At the end, data was split into 80\% training and 20\% testing subsets.

\subsection{Results}

\begin{figure}
    \centering
    \includegraphics[width=\linewidth]{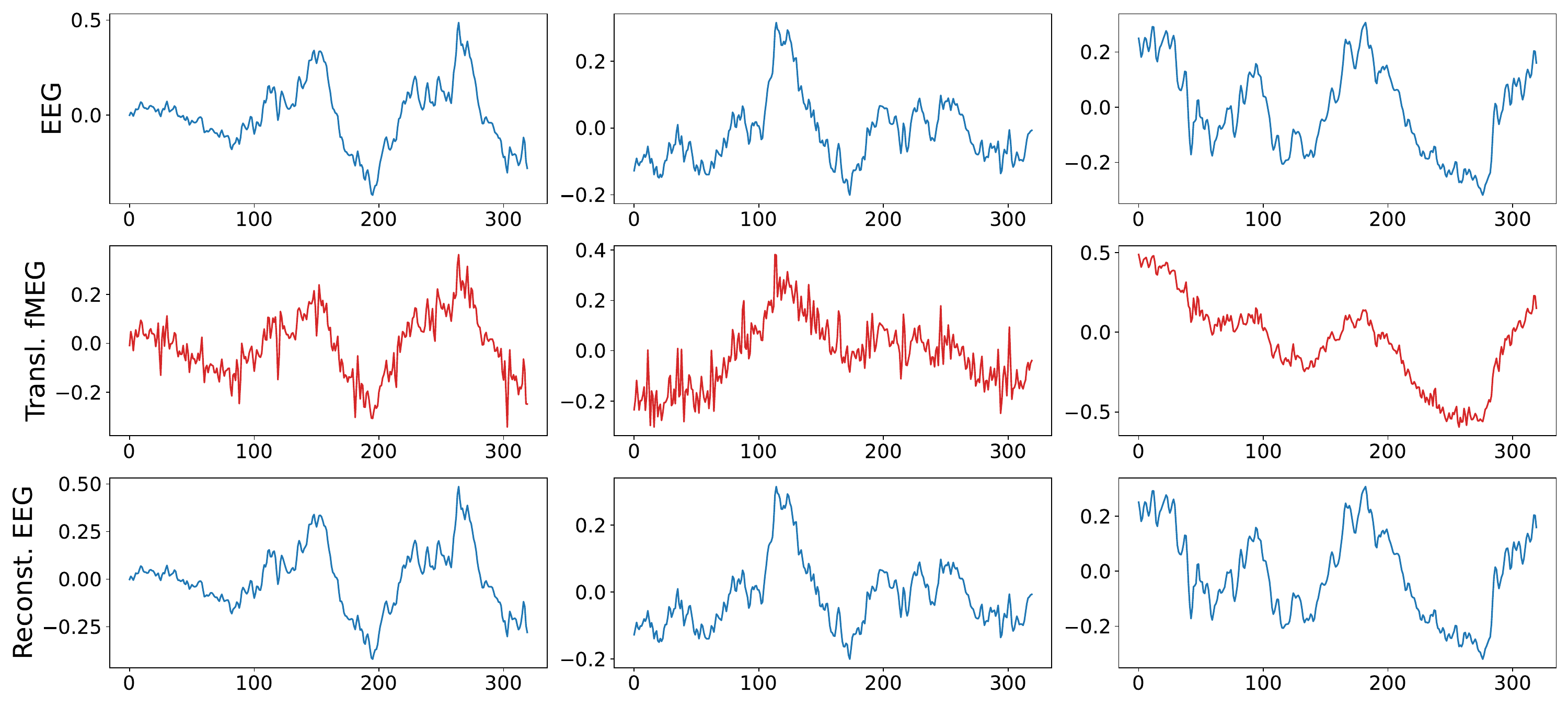}
    \caption{Examples of original EEG signals (top row) translated into fMEG signals (middle row) and then back into reconstructed EEG signals (bottom row) with our method. One can notice an excellent visual similarity between original and reconstructed EEG signals, although no cycle consistency regularization is applied.}
    \label{eeg_plot}
\end{figure}

\begin{figure}
    \centering
    \includegraphics[width=\linewidth]{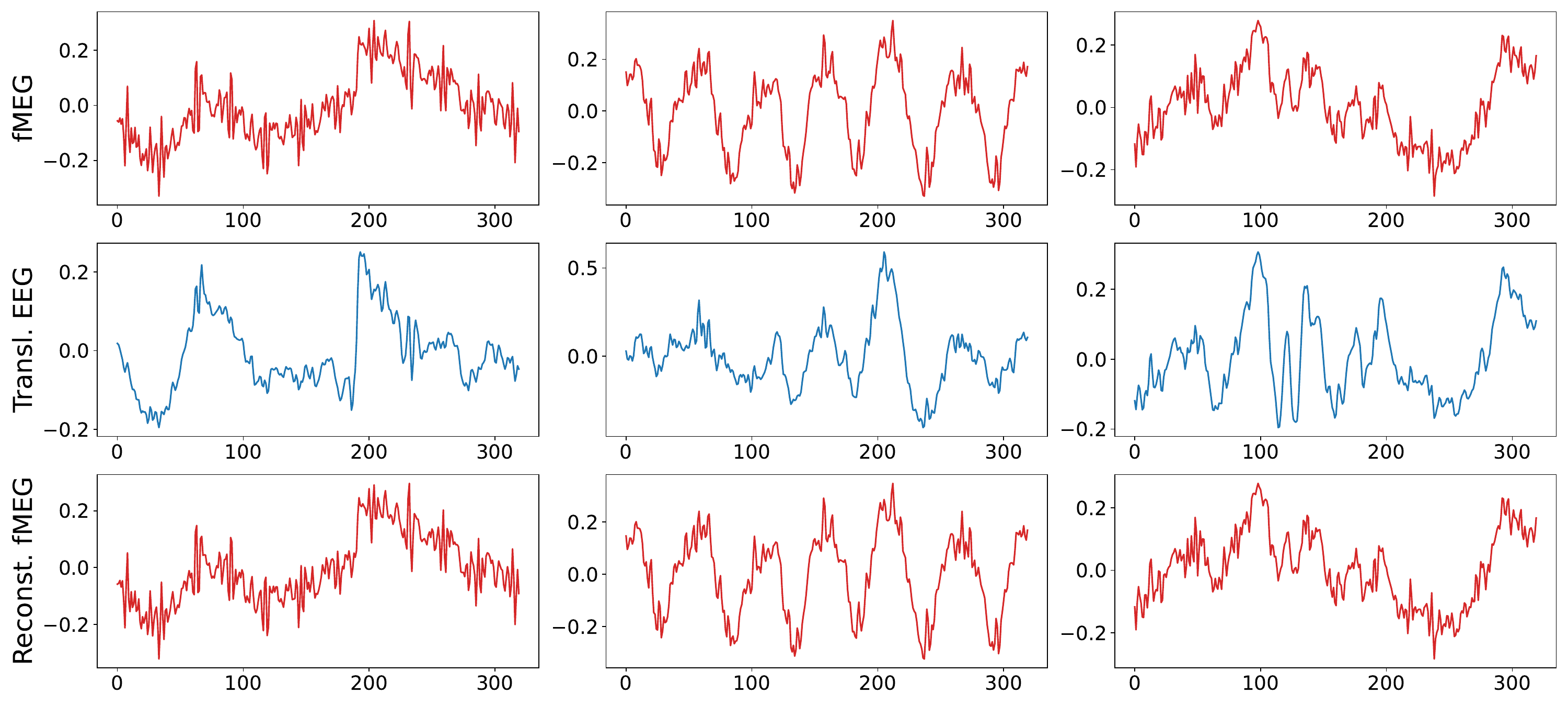}
    \caption{Examples of original fMEG signals (top row) translated into EEG signals (middle row) and then back into reconstructed fMEG signals (bottom row) with our method. As in Figure \ref{eeg_plot}, a visually perfect correlation between original and reconstructed signals can be observed.}
    \label{meg_plot}
\end{figure}

Using the previously trained models, we translate original EEG signals from the test dataset to their fMEG counterpart and then back to EEG (EEG - Translated fMEG - Reconstructed EEG), and we then do the same for the original fMEG signals (fMEG - Translated EEG - Reconstructed fMEG). Some examples of translated signals are shown in Figures \ref{eeg_plot} and \ref{meg_plot}.

We begin by analyzing the quality of reconstructed signals for both EEG and fMEG modalities, giving us good insights about the cycle consistent property of our method, and consequently if some information is lost during the translations. Ideally, reconstructed signals should perfectly match the original signals, as if we translate a sentence from one language to another and then back to the original language, the resulting sentence should not differ from the original one \cite{BTCCR_Brislin}. To this end, we compute the Mean Squared Error (MSE) between original and reconstructed signals, along with the ratio percentage between the MSE and the Mean Absolute Value (MAV). Results are summarized in Table \ref{mse_ratio_results}, where we compare ourselves to the state of the art (SOTA) for the EEG-fMEG translation problem, namely CycleGAN \cite{TSPN_Gallard}, and also to the original DDIB formulation \cite{DDIB_Su} as part of an ablation study to justify our modifications. As it can be noticed, our method greatly improves the values obtained with CycleGAN, by two orders of magnitude on the MSE, and by nearly 5\% on the MSE/MAV ratios, despite not relying on a cycle consistency loss. We also obtain better results than the original DDIB formulation, by one order of magnitude on the MSE, and by up to 0.6\% on the ratios, while having 4 times less passes through the model. We talk in more details about the Number of Function Evaluations (NFE) and discretization error in the Discussion section below.

\begin{table}
    \centering
    \resizebox{\columnwidth}{!}{
        \begin{tabular}{|c|c|cc|cc|cc|}
            \hline
            \multirow{2}{*}{\textbf{Data}} & \multirow{2}{*}{\textbf{MAV $(10^{-3})$}} & \multicolumn{2}{c|}{\textbf{CycleGAN} \cite{TSPN_Gallard} (NFE=1)} & \multicolumn{2}{c|}{\textbf{DDIB} (NFE=500)} & \multicolumn{2}{c|}{\textbf{Ours} (NFE=118)} \\ \cline{3-8} 
             &  & \multicolumn{1}{c|}{MSE $(10^{-3}) \downarrow$} & Ratio $(\%) \downarrow$ & \multicolumn{1}{c|}{MSE $(10^{-3}) \downarrow$} & Ratio $(\%) \downarrow$ & \multicolumn{1}{c|}{MSE $(10^{-3}) \downarrow$} & Ratio $(\%) \downarrow$ \\ \hline
            EEG & 129 $\pm$ 122 & \multicolumn{1}{c|}{6.96 $\pm$ 11.30} & 4.34 $\pm$ 4.43 & \multicolumn{1}{c|}{0.17 $\pm$ 0.40} & 0.14 $\pm$ 0.06 & \multicolumn{1}{c|}{\textbf{0.01 $\pm$ 0.06}} & \textbf{0.01 $\pm$ 0.01} \\ \hline
            fMEG & 143 $\pm$ 133 & \multicolumn{1}{c|}{9.04 $\pm$ 17.30} & 4.84 $\pm$ 5.69 & \multicolumn{1}{c|}{0.80 $\pm$ 1.98} & 0.67 $\pm$ 0.57 & \multicolumn{1}{c|}{\textbf{0.07 $\pm$ 0.61}} & \textbf{0.05 $\pm$ 0.20} \\ \hline
        \end{tabular}
    }
    \caption{Comparison with baseline models of the Mean Squared Error (MSE) between the original and reconstructed signals, and the ratio between the Mean Absolute Value (MAV) and MSE, for both EEG and fMEG modalities of the test dataset. Values are expressed as mean ± standard deviation. NFE stands for Number of Function Evaluations and indicates the number of passes through the model to translate one signal into its counterpart (e.g. EEG to translated fMEG).}
    \label{mse_ratio_results}
\end{table}

\begin{figure}
    \centering
    \begin{tikzpicture}[image/.style = {text width=0.45\textwidth, inner sep=0pt, outer sep=0pt}, node distance = 1mm and 1mm] 
        \node[image] (box1) {\includegraphics[width=\linewidth]{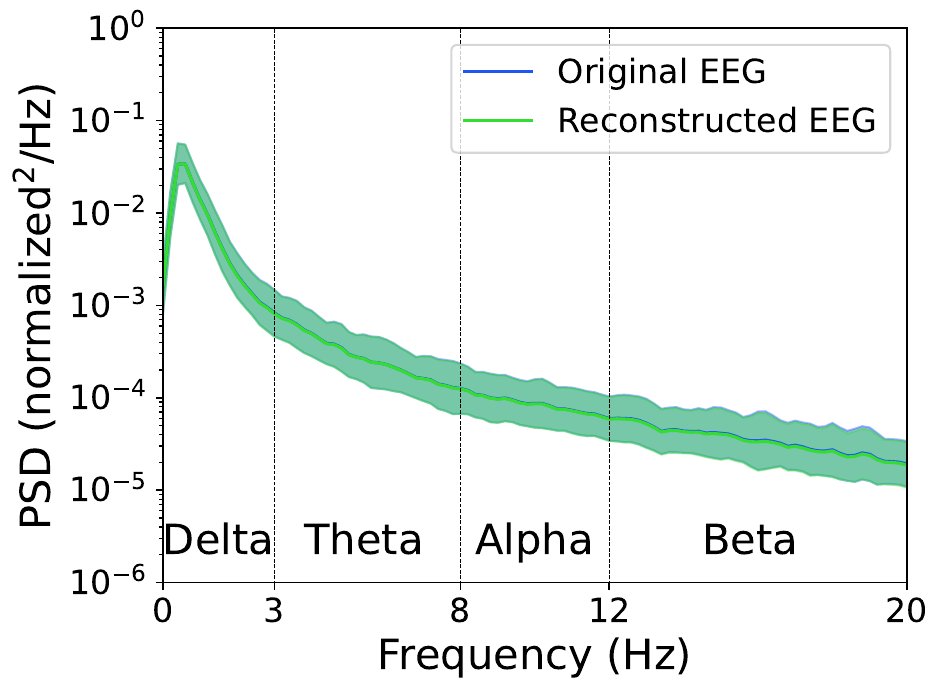}};
        \node[image, right = of box1] (box2) {\includegraphics[width=\linewidth]{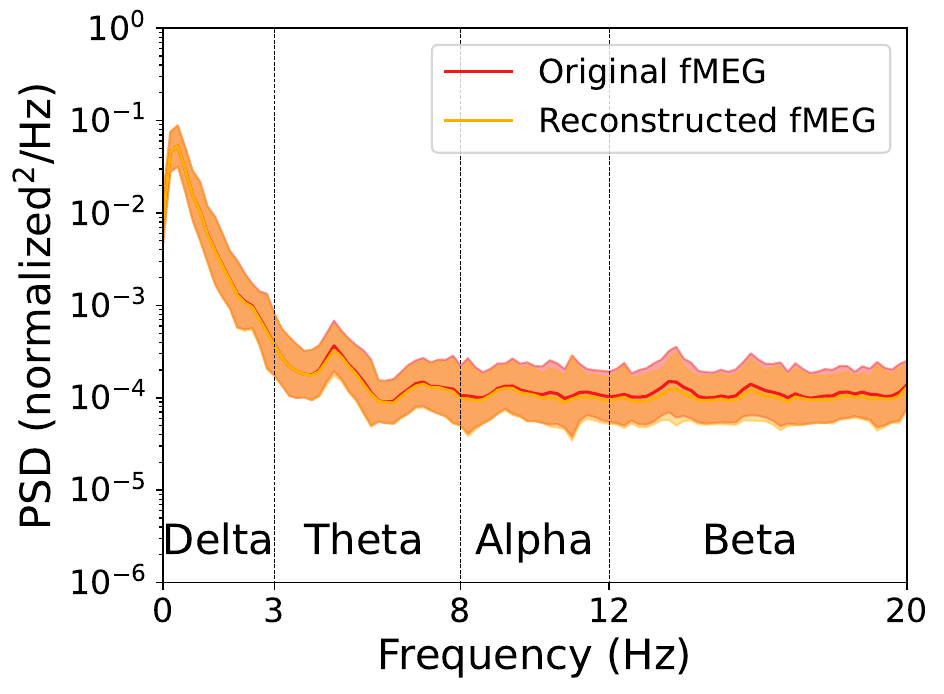}};
        \node[image, below = of box1] (box3) {\includegraphics[width=\linewidth]{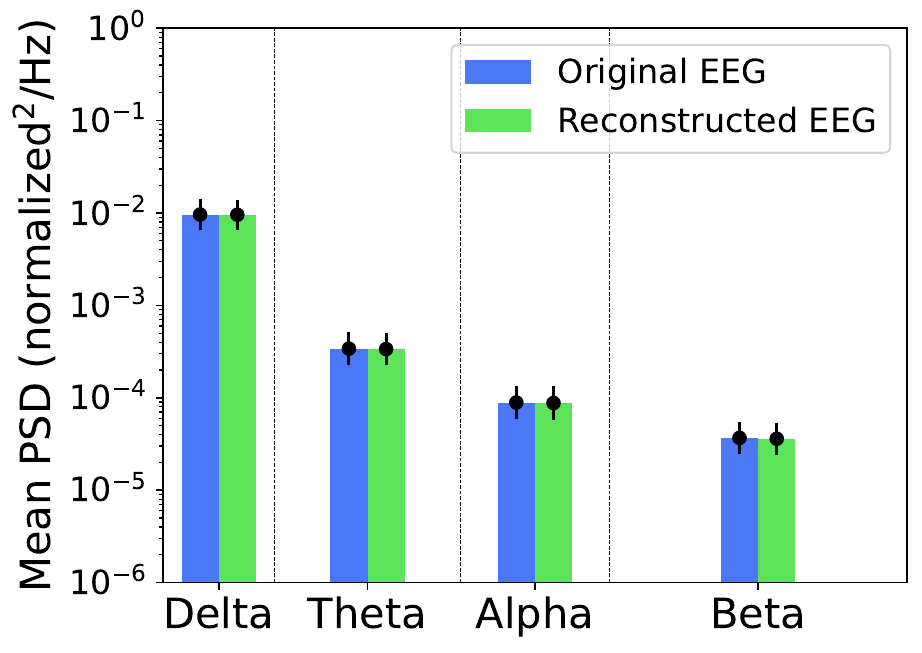}};
        \node[image, right = of box3] (box4) {\includegraphics[width=\linewidth]{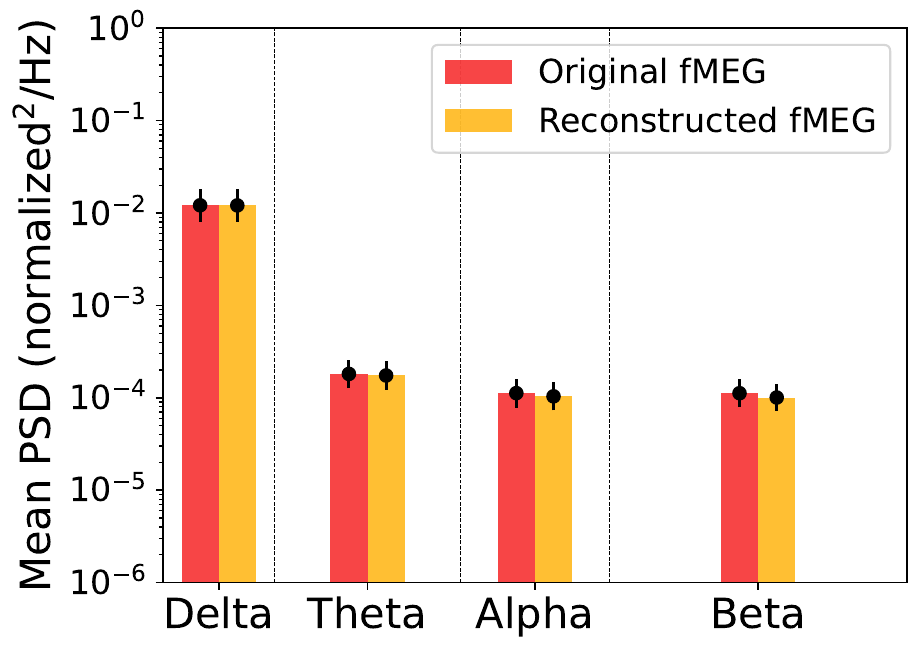}};
    \end{tikzpicture}
    \caption{Spectral power results with our method. Original signals spectrum compared to reconstructed signals spectrum for EEG (top left) and fMEG (top right), with the solid lines corresponding to the average and the shaded areas to the standard deviation. The mean spectral power over the frequency bands is also included for both EEG (bottom left) and fMEG (bottom right). Notice how the original and reconstructed spectrums match, as well as the mean per frequency band.}
    \label{psd_recon_plots}
\end{figure}

\begin{figure}
    \centering
    \begin{tikzpicture}[image/.style = {text width=0.45\textwidth, inner sep=0pt, outer sep=0pt}, node distance = 1mm and 1mm] 
        \node[image] (box1) {\includegraphics[width=\linewidth]{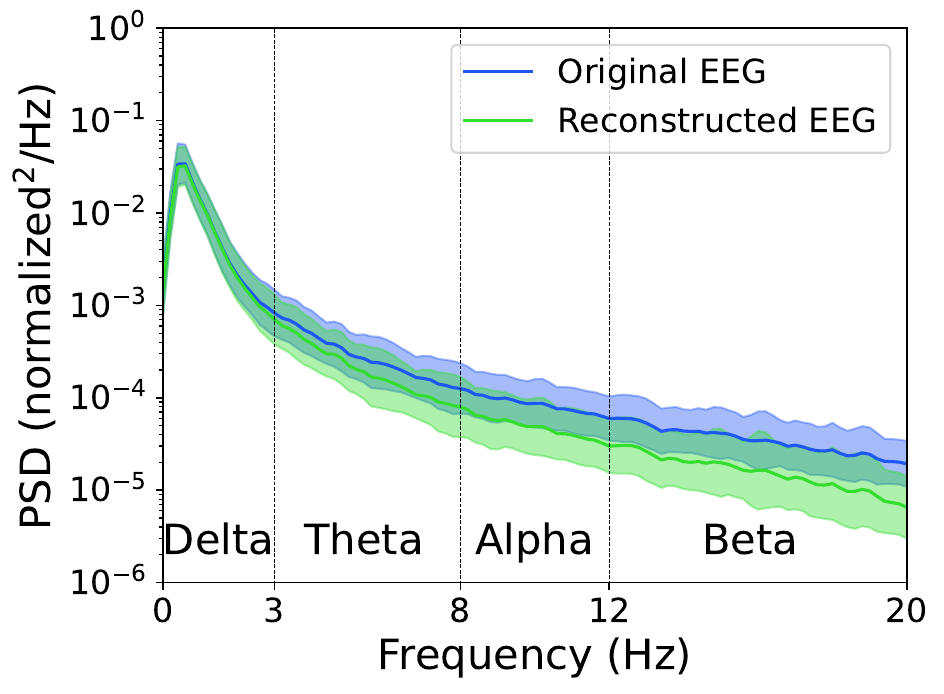}};
        \node[image, right = of box1] (box2) {\includegraphics[width=\linewidth]{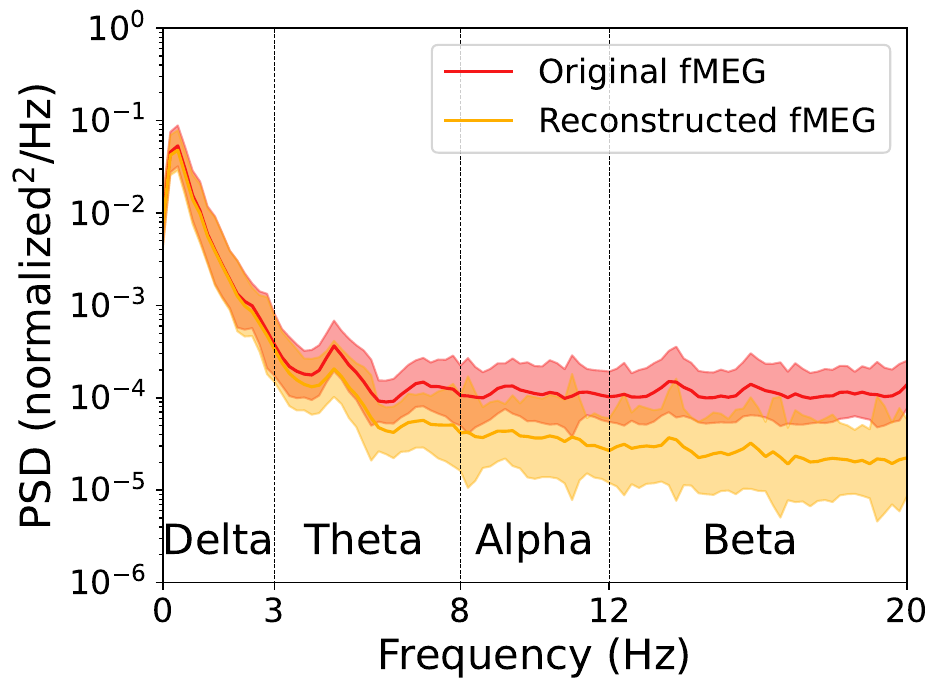}};
        \node[image, below = of box1] (box3) {\includegraphics[width=\linewidth]{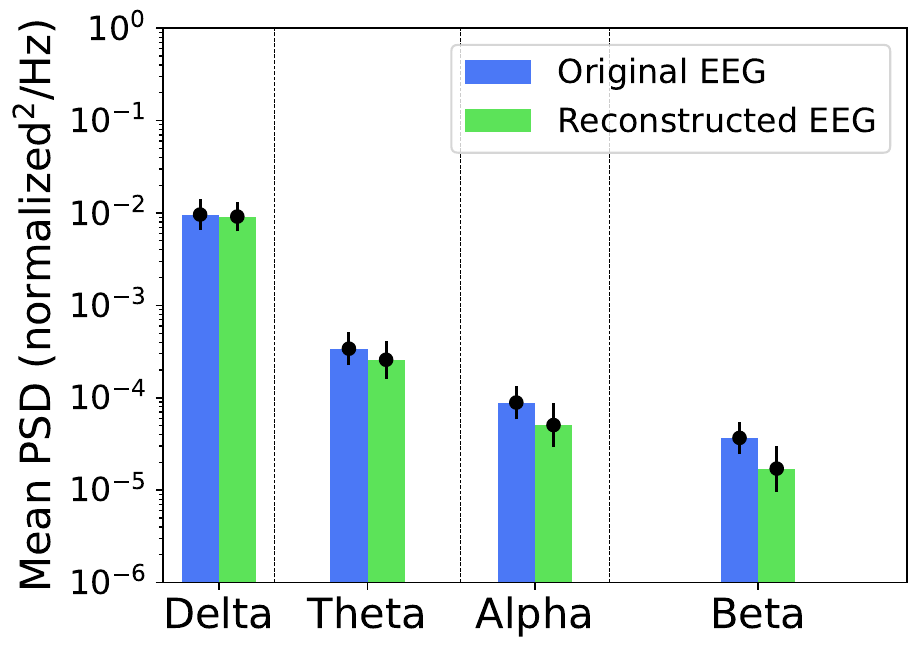}};
        \node[image, right = of box3] (box4) {\includegraphics[width=\linewidth]{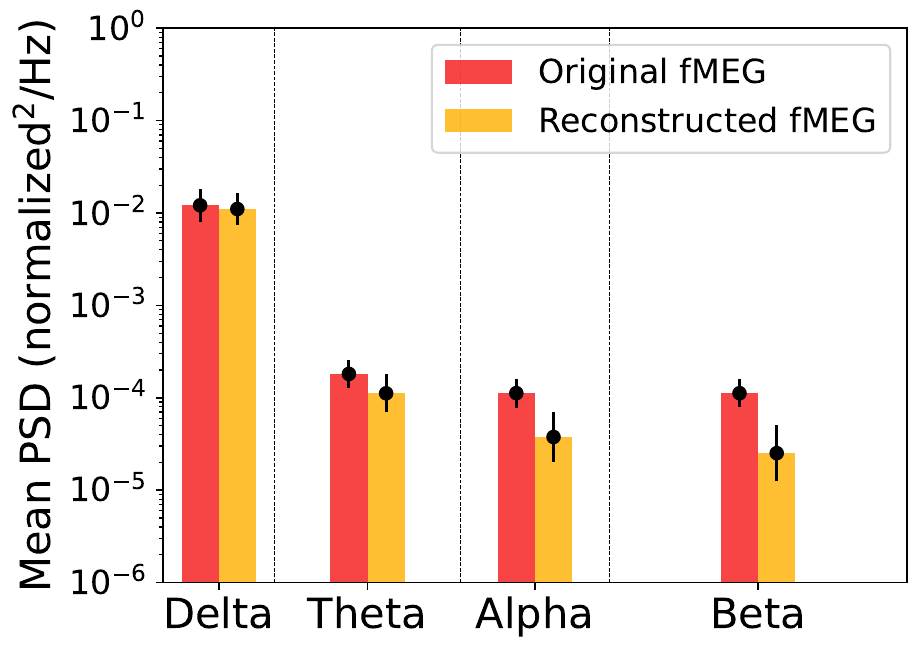}};
    \end{tikzpicture}
    \caption{Spectral power results with DDIB. Original signals spectrum compared to reconstructed signals spectrum for EEG (top left) and fMEG (top right), with the solid lines corresponding to the average and the shaded areas to the standard deviation. The mean spectral power over the frequency bands is also included for both EEG (bottom left) and fMEG (bottom right). One can observe a slight shift starting from the Theta band.}
    \label{psd_recon_plots_ddib}
\end{figure}

In addition to the time domain analysis, we also provide a comparison in the frequency domain to assess the quality of reconstructed signals. Thereby, we compute the power spectral density (PSD) by averaging spectrums of each signal obtained using a discrete Fourier transform. We then split the resulting spectrum into four frequency bands, namely Delta (0.5-3 Hz), Theta (3-8 Hz), Alpha (8-12 Hz) and lower Beta (12-20 Hz), and compute the average value of each of these bands. Results from our method can be found in Figure \ref{psd_recon_plots}, and we also provide results obtained with the original DDIB framework in Figure \ref{psd_recon_plots_ddib}. As in the time domain, we hugely improve previous results from the SOTA for the EEG-fMEG problem. Unlike \cite{TSPN_Gallard} where important divergences could be noticed for the spectrums and the mean spectral power for both signal types, here we manage to obtain a near perfect correlation between the original and reconstructed signals. This analysis also highlights an important difference that could not be inferred with the MSE in the time domain, as a small shift can be observed with the original DDIB in Figure \ref{psd_recon_plots_ddib}, resulting in a partial loss of information in the higher frequency bands. Notice that this shift is not present with our improvements, more on this in the Discussion section.

\begin{figure}
    \centering
    \begin{tikzpicture}[image/.style = {text width=0.45\textwidth, inner sep=0pt, outer sep=0pt}, node distance = 1mm and 1mm] 
        \node[image] (box1) {\includegraphics[width=\linewidth]{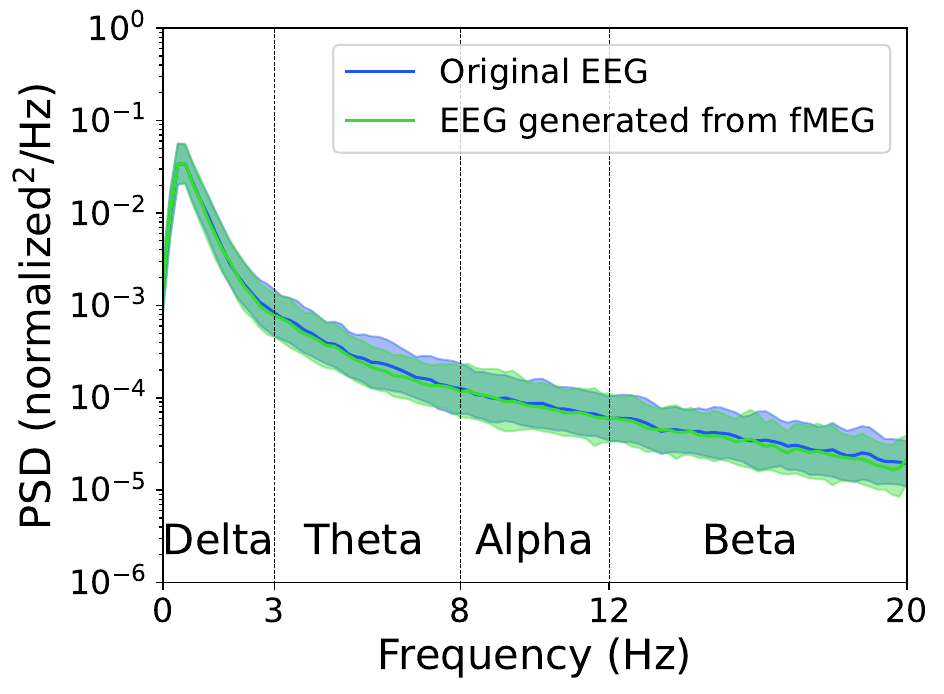}};
        \node[image, right = of box1] (box2) {\includegraphics[width=\linewidth]{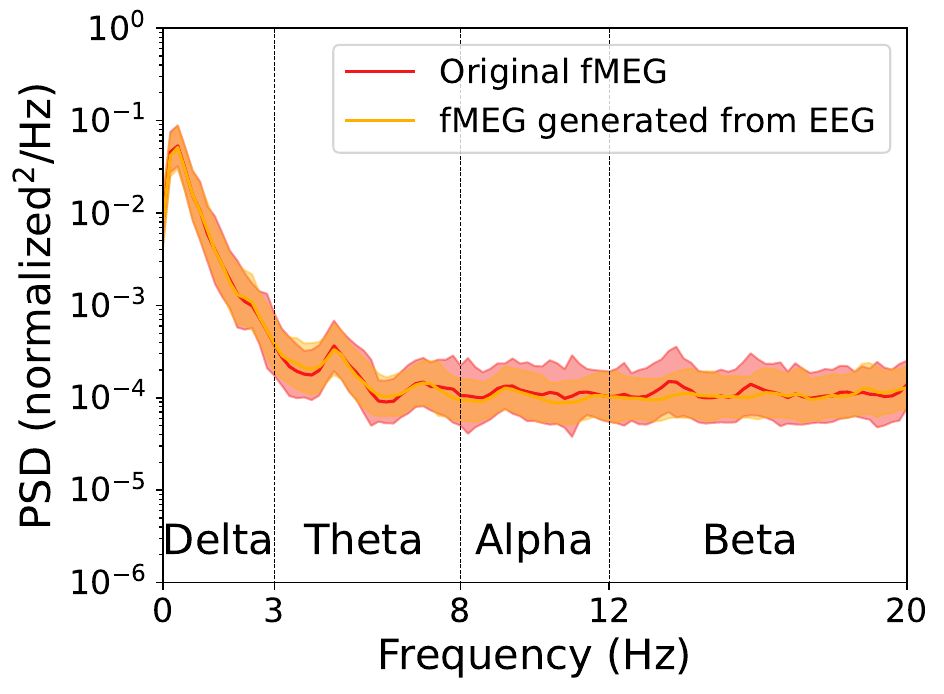}};
        \node[image, below = of box1] (box3) {\includegraphics[width=\linewidth]{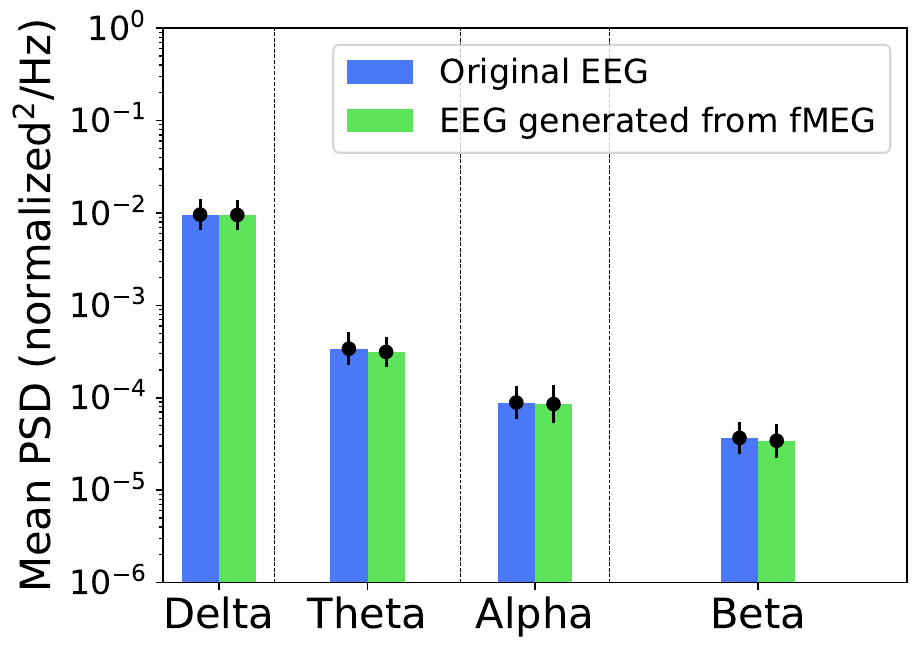}};
        \node[image, right = of box3] (box4) {\includegraphics[width=\linewidth]{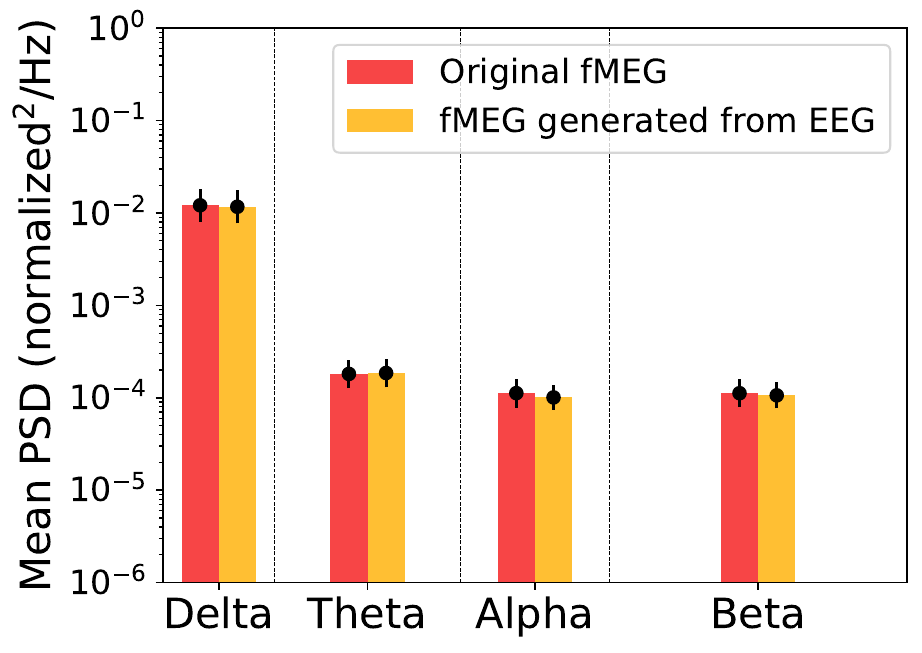}};
    \end{tikzpicture}
    \caption{Spectral power results with our method. Original signals spectrum compared to translated signals spectrum for EEG (top left) and fMEG (top right), with the solid lines corresponding to the average and the shaded areas to the standard deviation. The mean spectral power over the frequency bands is also included for both EEG (bottom left) and fMEG (bottom right). As in Fig.\ref{psd_recon_plots}, notice how the original and translated spectrums match, as well as the mean per frequency band.}
    \label{psd_plots}
\end{figure}

\begin{figure}
    \centering
    \begin{tikzpicture}[image/.style = {text width=0.45\textwidth, inner sep=0pt, outer sep=0pt}, node distance = 1mm and 1mm] 
        \node[image] (box1) {\includegraphics[width=\linewidth]{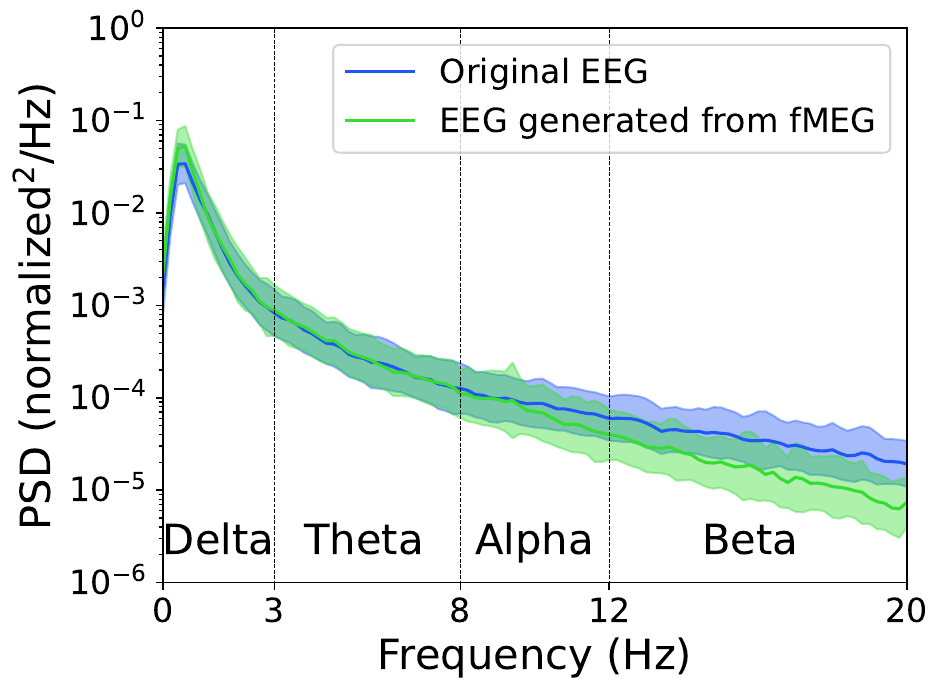}};
        \node[image, right = of box1] (box2) {\includegraphics[width=\linewidth]{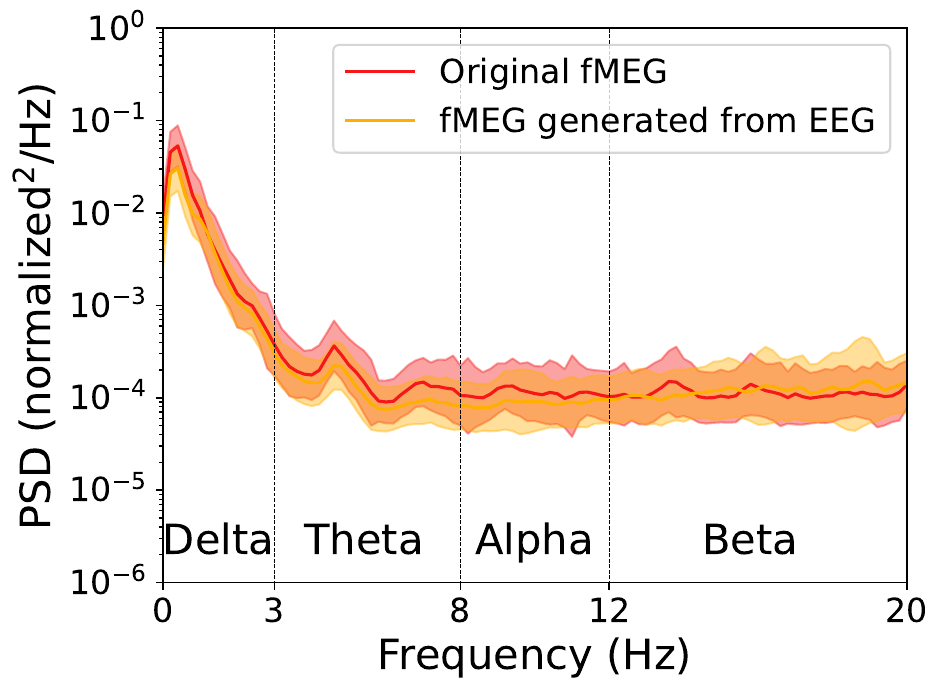}};
        \node[image, below = of box1] (box3) {\includegraphics[width=\linewidth]{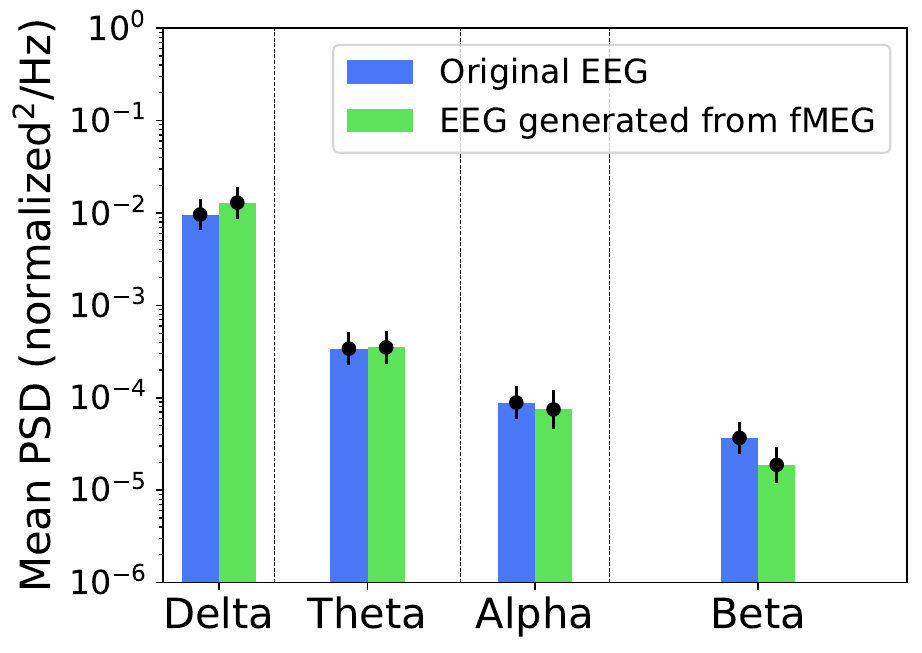}};
        \node[image, right = of box3] (box4) {\includegraphics[width=\linewidth]{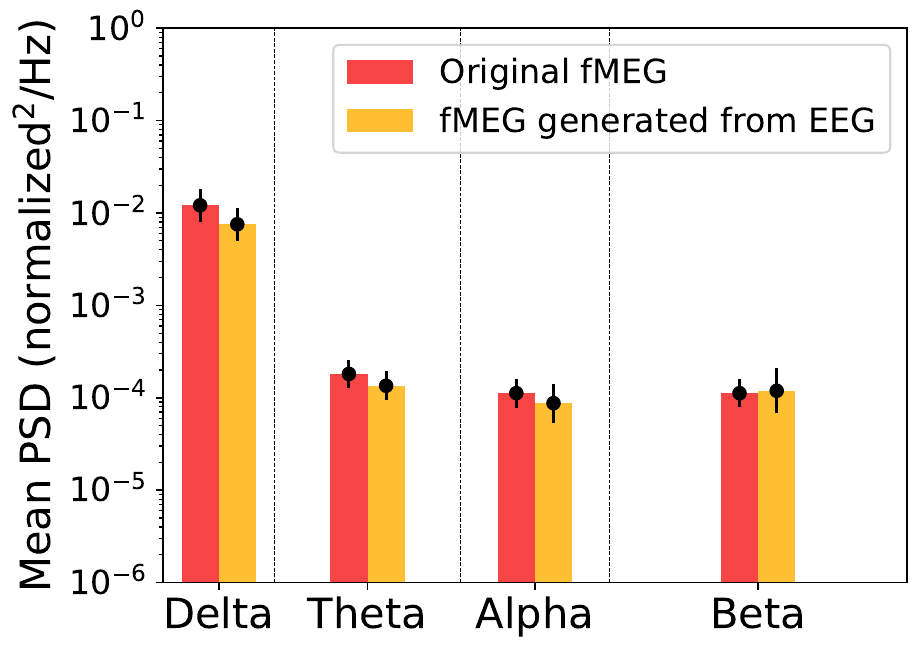}};
    \end{tikzpicture}
    \caption{Spectral power results with DDIB. Original signals spectrum compared to translated signals spectrum for EEG (top left) and fMEG (top right), with the solid lines corresponding to the average and the shaded areas to the standard deviation. The mean spectral power over the frequency bands is also included for both EEG (bottom left) and fMEG (bottom right).}
    \label{psd_plots_ddib}
\end{figure}

We then seek to compare the quality between original and translated signals, in other words between original EEG and EEG translated from fMEG as well as between original fMEG and fMEG translated from EEG. Since we don't have any real pair of EEG and fMEG data, we are only able to study if the frequency properties are consistent, thus we once again compute the power spectral density. One should note that this analysis is as important as the one between original and reconstructed signals. This indeed allows to make sure that the models don't converge to wrong maps like the identity, in which case e.g. an original EEG and a reconstructed EEG signal would perfectly match because the translated fMEG would still in fact be the original EEG. In this way, we can ensure that the models are learning meaningful data features. The plot results are presented in Figure \ref{psd_plots} for our method, and Figure \ref{psd_plots_ddib} for the original DDIB formulation. We achieve excellent coverage across all frequency bands in the EEG and fMEG cases, whereas the original DDIB version suffers once again from a slight shift, especially noticeable with EEG. In both cases, the outcome is far more accurate than the SOTA CycleGAN \cite{TSPN_Gallard} (only showed for fMEG), where discrepancies appear in all but the Delta band.

\section{Discussion}

\subsection{Biomarkers transfer}

In this paper, we focused on providing a method that could translate EEG and fMEG complex neurophysiological signals while making sure time and frequency properties of both types are respected. From this work, we hope to facilitate the characterization of brain activity patterns of the fetus. Whilst this analysis could be the subject of a separate paper, we seek to provide a first investigation on the transfer of well-known EEG neurobiomarkers to fMEG.

\begin{figure}
    \centering
    \includegraphics[width=\linewidth]{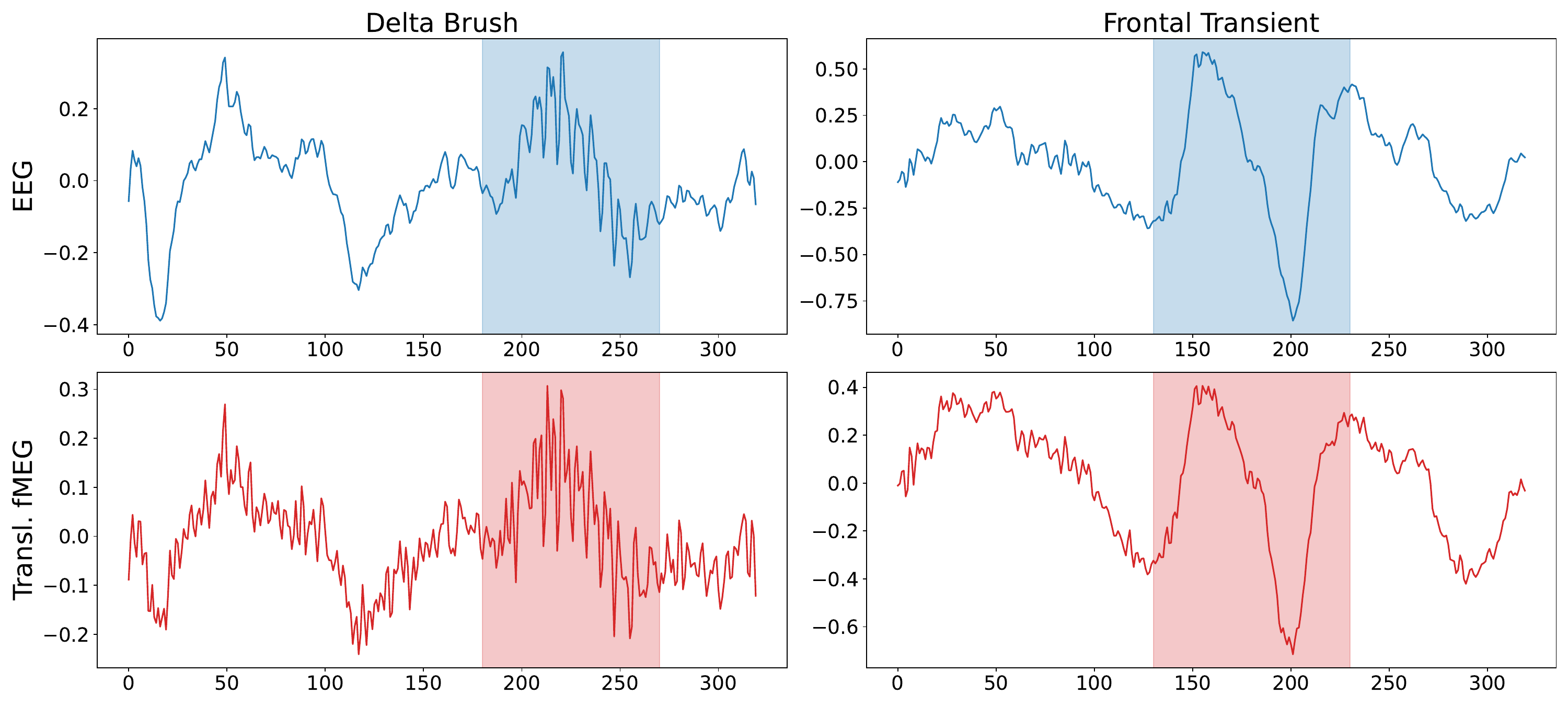}
    \caption{Example of a sample containing a delta brush (left column) and frontal transient (right column) translated from EEG (top row) to fMEG (bottom row).}
    \label{eeg_db_ft_plot}
\end{figure}

As discussed in the Introduction section, early brain activity is marked by patterns indicative of brain development status and possible neurological issues. Among these, and given the gestational ages of participants in our dataset, we are able to analyze the transfer of 2 specific activities, namely delta brushes (DB) and frontal transients (FT). As in \cite{TSPN_Gallard}, we only inspect the results visually over a small number of events, though signals frequency properties are already assessed with Figure \ref{psd_plots}. An example of transfer from EEG to fMEG of a sample containing a DB and a sample containing a FT is given in Figure \ref{eeg_db_ft_plot}. As it can be observed, the fMEG-translated DB still consists of a slow wave within which faster activities are nested. This is a different result from CycleGAN \cite{TSPN_Gallard} where the fast oscillations were in fact soften. Concerning the FT in the fMEG domain, we obtain a result similar to CycleGAN with the slow oscillation replicated from the original EEG, and with the global appearance of small high frequencies variations.

\subsection{Function evaluations and discretization error}

As indicated in Table \ref{mse_ratio_results} with the number of function evaluations (NFE), CycleGAN only needs 1 pass through the model to translate a signal into its counterpart, while we use respectively 500 and 118 passes in total for DDIB and our method. This is due to the iterative nature of diffusion models progressively adding and removing noise at each step, making them fall behind GANs in terms of efficiency. 

If we now focus only on DDIB and our method, one can notice that our improvements lead to better results while having 4 times less NFE. The most prominent problem appearing with DDIB is the high frequency shift that can be observed in Figures \ref{psd_recon_plots_ddib} and \ref{psd_plots_ddib}. Although we can draw a parallel between this and the PF-ODE discretization error, as the error tends to get worse as we reduce the total number of steps, we also observed that increasing the steps, while keeping reasonable numbers, would not really mitigate the problem. We hypothesize that by being restrained to a first-order solver, the original DDIB framework is, to a certain extent, limited in its ability to preserve fine details, like the high frequencies of our EEG and fMEG signals. On the opposite side, by relying on Heun's 2nd order method, our modifications allow us to not be affected by this problem, while having a good trade-off between the discretization error and sampling speed.

As a side note, distillation could be a versatile solution usable in most diffusion model cases to reduce the total number of steps, at a cost of a more complex training pipeline \cite{PDFSDM_Salimans, IBCD_Lee}.

\subsection{Limitations and possible enhancements}

Our study on 1D unpaired diffusion translation models extended beyond DDIB and our proposed improved version. We explored original ideas as well as works derived from other existing methods like CycleNet \cite{CycleNet_Xu}. But a consistent observation we noticed across these experiments is that unpaired translation models relying on a multi-component loss (similar to CycleGAN \cite{CycleGAN_Zhu} which combines a loss part for data fidelity and another loss part for cycle-consistency) tend to be difficult to train with low-dimensional data. Models indeed struggle to balance the loss components, resulting either in a overemphasis on signal realism at the expense of coherent domain translation mappings, or simply collapse into learning the identity mapping.

Regarding the evaluation and choice of the best trained diffusion models, we considered using the Context-FID metric \cite{PSAGAN_Jeha}, which is a Frechet Inception Distance (FID)-like score for time series. As in the original FID metric \cite{FID_Heusel} used in image generation, Context-FID computes the distance between the distributions of real and generated data in the feature space of a pretrained network, more specifically a 1D Inception-style model trained on time series data. While this offers a convenient and automated way to assess the realism and diversity of generated data, FID-based methods require a sufficiently large number of samples (at least 10.000 if we refer to the literature) to yield meaningful estimates. Since our datasets didn't meet this scale, we indeed observed that using Context-FID in our setting led to unstable evaluations that were not reliable for model selection. As a result, we instead based our choice on the diffusion loss and complemented this with a manual assessment of the power spectral density quality to ensure that the properties of the generated signals are correct.

We also want to mention that in the absence of EEG-fMEG paired data, it remains difficult to draw any firm conclusions about the correctness of the translated signals. We can still mention the possibility of using MEG with premature infants, and therefore in the future of having paired EEG - MEG data which could potentially help to improve and consolidate our work, although MEG in premature babies and fetal MEG have divergences related to data preprocessing and in-utero/ex-utero environment.

Another possibility for future improvement would be to use the multi-channel nature of the data, which we are not doing in our case for the sake of simplification given that we are among the first to explore this problem. Relevant information that could be used to make training more robust is certainly hidden in the interactions between the different channels.

\section{Conclusion}

In this paper, we proposed an unpaired diffusion translation model for premature EEG and fetal MEG neurophysiological signals, based on dual diffusion bridges. We showed through extensive experimentation that our method produces results that are much more consistent and accurate than the state-of-the-art for this problem, namely CycleGAN, counteracting the mode collapse problem in the process. We also demonstrated in an ablation study that, thanks to our diffusion and numerical integration modifications, we improve the results obtained with the original DDIB framework while reducing computational costs.

Overall, our new state-of-the-art model for premature EEG and fetal MEG translation holds promise for advancing our understanding of early brain activity and potentially improving diagnostic tools. We believe that our method can also serve as a basis for other unpaired signal translation applications in a wide variety of fields.

\section*{Glossary}

\begin{itemize}
    \item Cycle consistency regularization: Ensures that translating a sample to another domain and back returns the original sample, promoting meaningful and reversible transformations
    \item Markov chain: Stochastic process where the probability of an event depends solely on the previous state
    \item Mode collapse: Failure mode observed in Generative Adversarial Networks, where the model produces outputs that are less diverse than expected
    \item ODE/SDE solver: Numerical method used to approximate solutions to ordinary or stochastic differential equations over time
    \item Schrödinger bridges: Most likely stochastic path connecting two probability distributions
    \item Wiener process: Continuous-time stochastic process with independent Gaussian increments, often used to model Brownian motion
\end{itemize}

\section*{Acronyms}

\begin{itemize}
    \item DB: Delta brushes
    \item DDIB: Dual diffusion implicit bridges
    \item DDIM: Denoising diffusion implicit models
    \item DDPM: Denoising diffusion probabilistic models
    \item EDM: Elucidated diffusion models
    \item EEG: Electroencephalography
    \item FID: Frechet Inception distance
    \item fMEG: Fetal magnetoencephalography
    \item FT: Frontal transients
    \item GAN: Generative adversarial networks
    \item GB: Gigabyte
    \item GPU: Graphics processing unit
    \item IBI: Inter-burst intervals
    \item MAV: Mean absolute value
    \item MSE: Mean squared error
    \item NFE: Number of function evaluations
    \item NLEO: Nonlinear energy operator
    \item ODE: Ordinary differential equation
    \item PF-ODE: Probability flow ordinary differential equation
    \item PSD: Power spectral density
    \item SDE: Stochastic differential equation
    \item SOTA: State of the art
    \item TTA-SW: Theta temporal activity in coalescence with slow-waves
    \item VRAM: Video random access memory
    \item wGA: Weeks of gestational age
\end{itemize}

\section*{Acknowledgments}

This work was supported by joint funding from the Agence Nationale de la Recherche (ANR-21-CE19–0046 fMEG-OPM and ANR VIVAH) and the Deutsche Forschungsgemeinschaft (PR 496/11–1). This work was also supported by grants from the Région Hauts-de-France and the VIVAH project. This work was partially supported by the Helmholtz Association (Helmholtz Excellence Networks: EXNET-01-17). This work was performed using HPC resources from GENCI–IDRIS (Grant 2023-AD011014435). The authors are grateful to the Amiens University Hospital EEG technicians and Tübingen University Hospital MEG
technicians for data acquisition. We also thank the families who consented to take part in the study.

\section*{Ethical approval}

The EEG study was approved by the local CPP Nord-Ouest I ethics committee, and one or both parents were informed about the study and provided their written informed consent (ID-RCB: 2021-A02556–35). The fMEG study was approved by the local ethics committee of the Medical Faculty of the University of Tübingen, and the consent to participate was signed by the mother (511/2015BO1 and 330/2010BO1).

\bibliography{refs}

\newpage

\setcounter{affn}{0}

\begin{frontmatter}

\title{Diffusion-based translation between unpaired spontaneous premature neonatal EEG and fetal MEG \\ \large Supplementary Material}

\begin{abstract}
The following pages contain the supplementary material for the ``Diffusion-based translation between unpaired spontaneous premature neonatal EEG and fetal MEG''
\end{abstract}

\end{frontmatter}

\section{Dataset}

\subsection{Population}

The distribution of participants per signal modality can be found in Table \ref{participants_distribution}. All infants of the EEG study had normal birth parameters, an APGAR score above 5 at 5 minutes, a normal clinical neurological assessment, and no risk factors for brain damage, with EEG evaluations confirming brain maturation corresponding to their gestational age. EEG assessment as part of the follow-up when leaving the neonatal intensive care unit had to meet the normal standards set by the EEG monitoring guidelines of the French Society of Clinical Neuroscience.

MEG study participants were recruited from Tübingen and its surrounding areas, with mothers having uncomplicated singleton pregnancies.

\begin{table}[h!]
    \centering
    \begin{tabular}{|l|llll|l|}
        \hline
        \multirow{2}{*}{\textbf{Signal type}} & \multicolumn{4}{l|}{\textbf{wGA}} & \multirow{2}{*}{\textbf{Total}} \\
        \cline{2-5} & \multicolumn{1}{l|}{\textbf{34}} & \multicolumn{1}{l|}{\textbf{35}} & \multicolumn{1}{l|}{\textbf{36}} & \textbf{37} & \\
        \hline
        EEG & \multicolumn{1}{l|}{7} & \multicolumn{1}{l|}{9} & \multicolumn{1}{l|}{7} & 7 & 30 \\
        \hline
        fMEG & \multicolumn{1}{l|}{13} & \multicolumn{1}{l|}{11} & \multicolumn{1}{l|}{12} & 8 & 44 \\
        \hline
    \end{tabular}
    \caption{Number of participants per signal type according to their gestational age at the recording date.}
    \label{participants_distribution}
\end{table}

\subsection{Acquisition}

High-resolution EEG recordings of sleeping premature neonates were performed in the incubator, using a 64 or 124-channel HydroCel Geodesic Sensor Net with an Electrical Geodesic NetAmps 200 amplifier passing a DC–50 Hz filtered digitized signal to the Electrical Geodesics Net Station software (Version 5). EEG was digitized at a 1000 Hz sampling rate, with a Cz vertex electrode as the reference, while maintaining the electrode impedance below 5 k$\Omega$.

fMEG recordings were performed using a SARA (SQUID Array for Reproductive Assessment, VSM MedTech Ltd., Port Coquitlam, Canada) system installed at the University of Tübingen fMEG Center, and designed to adapt the established MEG technique to the specific requirements of fetal measurements. This system is composed of 156 primary magnetic sensors distributed over a concave array shaped to match the form of the gravid abdomen, and also includes 29 reference sensors. fMEG was recorded at a sampling rate of 610.352 Hz, in a magnetically shielded room (Vakuumschmelze, Hanau, Germany) to reduce external magnetic interference.

\subsection{Preprocessing}

All the EEG and fMEG pre-processing steps were carried out using MNE 1.2.1 and Python 3.10.6.

\begin{figure}
    \centering
    \includegraphics[width=0.8\linewidth]{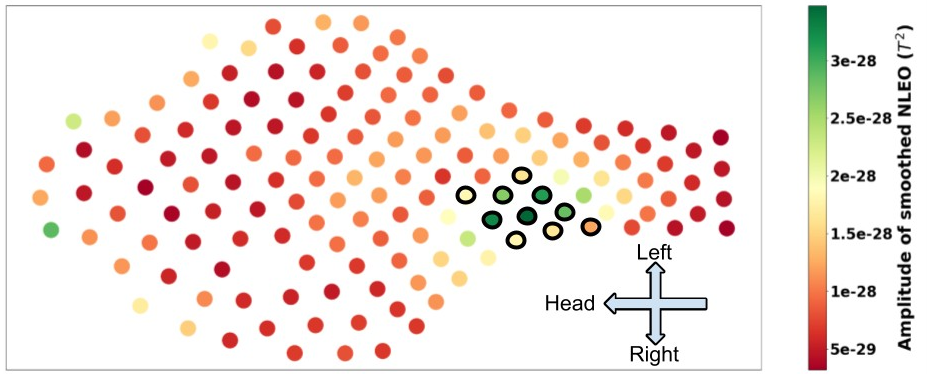}
    \caption{Spatial amplitude distribution of smoothed NLEO for one participant, with selected region of interest shown with black circles. The head of the mother is located on the left of the figure, her left on the top and her right on the bottom. In this case, the centroid is localized around the center of the lower abdomen of the mother.}
    \label{NLEO_SA_distrib}
\end{figure}

The 10 EEG channels chosen for detecting spontaneous neural activity and spanning the entire head are the following: Fp2-C4, Fp1-C3, C4-O2, C3-O1, Fp2-T4, Fp1-T3, T4-O2, T3-O1, Fz-Cz, and Cz-Pz. For each sample $i$ of the signal, the NLEO index was calculated as:
\begin{equation}
    NLEO(x) = x(i)x(i-3) - x(i-1)x(i-2)
\end{equation}
We illustrate in Fig. \ref{NLEO_SA_distrib} an example of the spatial distribution of the NLEO mean amplitude over the channels. The normalization of range [-1,1] is done for both datasets (EEG and fMEG) using the following formula:
\begin{equation}
    x_{norm} = 2 \frac{x - min(x)}{max(x) - min(x)} - 1
\end{equation}
where $x$ is the signal amplitude before normalization, and $min(x)$ and $max(x)$ are the minimum and maximum amplitude values of the whole dataset.

Final subsets distribution is summarized in Table \ref{subsets_distribution}.

\begin{table}
    \centering
    \begin{tabular}{|l|l|l|l|}
        \hline
        \textbf{Signal type} & \textbf{Train} & \textbf{Test} & \textbf{Total} \\ \hline
        EEG                  & 8542           & 2090          & 10632          \\ \hline
        fMEG                 & 7684           & 1873          & 9557           \\ \hline
    \end{tabular}
    \caption{Number of EEG and fMEG burst segments per subset.}
    \label{subsets_distribution}
\end{table}

\end{document}